\documentclass[twocolumn,showpacs,preprintnumbers,amsmath,amssymb]{revtex4}
\usepackage{graphicx}
\usepackage{epstopdf}
\usepackage{amssymb}
\usepackage{dcolumn}
\usepackage{bm}

\begin{document}


\title{Phase transitions in Pareto optimal complex networks}

\author{Lu\'is F. Seoane$^{1,2}$ and Ricard Sol\'e$^{1,2,3}$ }
\affiliation{  
    \\ $^1$ ICREA-Complex  Systems   Lab,  Universitat Pompeu Fabra  -  PRBB,
    Dr. Aiguader 88, 08003  Barcelona, Spain
    \\ $^2$ Institut de Biologia Evolutiva, UPF-CSIC, Passg Barceloneta, 08003 Barcelona
    \\$^3$ Santa Fe Institute, 1399 Hyde Park Road, New Mexico 87501, USA
  }

\begin{abstract}

  The organization of interactions in complex systems can be described by networks connecting
different units. These graphs are useful representations of the local and global complexity of the
underlying systems. The origin of their topological structure can be diverse, resulting from
different mechanisms including multiplicative processes and optimization. In spatial networks or in
graphs where cost constraints are at work, as it occurs in a plethora of situations from power grids
to the wiring of neurons in the brain, optimization plays an important part in shaping their
organization. In this paper we study network designs resulting from a Pareto optimization process,
where different simultaneous constraints are the targets of selection. We analyze three variations
on a problem finding phase transitions of different kinds. Distinct phases are associated to
different arrangements of the connections; but the need of drastic topological changes does not
determine the presence, nor the nature of the phase transitions encountered. Instead, the functions
under optimization do play a determinant role. This reinforces the view that phase transitions do
not arise from intrinsic properties of a system alone, but from the interplay of that system with
its external constraints.

\end{abstract}

\pacs{64.60.aq, 64.60.Bd, 64.60.-i, 87.55.de} 

\maketitle

\section{Introduction}
	\label{sec:1}

  Optimization is a key goal in engineered systems and is traditionally assumed to be part of the
intrinsic dynamics of natural evolving systems \cite{Dawkins1997}. The engineering perspective,
associated to man-made objects and structures, is specially obvious when dealing with large-scale,
interconnected units, as it occurs in very large integrated circuit design \cite{Ozaktas1992,
Chen1999, BassettBullmore2010} or spatially-extended infrastructures such as power grids
\cite{WattsStrogatz1998, AmaralStanley2000, Barthelemy2011, RosasCasalsSole2014} and transportation
or distribution networks \cite{AmaralStanley2000, Barthelemy2011, Kansky1963, Pitts1965,
SenManna2003, GastnerNewman2006, CarvalhoArrowsmith2009}. In these cases, engineers cope with
interfering constraints related to materials, space, packing, wiring, or dissipation costs. The
staggering complexity of these designed systems can be addressed by algorithms that deal with
multidimensional problems.

  In biological systems, important network topologies have been shown to result from optimality
\cite{CluneLipson2013, MengitsuClune2015}. These include transportation networks in living organisms
\cite{Murray1926, WestEnquist1997, WestEnquist1999, BanavarRinaldo1999, GafiychukLubashevsky2001}
where  optimization is reached by means of fractal trees that guarantee a low cost and efficient
location of resources. Similarly, neural circuits display optimal features over a wide range of
scales \cite{BassettBullmore2010, CuntzSegev2007, PerezEscuderoPolavieja2007,
HasenstaubSejnowski2010, CuntzHausser2010, WedeenTseng2012}. The packing and interconnectivity in
some cortical areas seem compatible with design principles shared by high-density electronic designs
\cite{BassettBullmore2010}.

  In all the previous examples tradeoffs between efficiency and cost are present. Packing many
components in a given spatial domain is desirable because of cost minimization of connections, but
dissipation of energy or wiring constraints will also be at work. What kinds of topologies result
when considering multiple constraints? This problem has been addressed  by explicitly introducing
efficiency measures $E$ (such as average path length) along with cost constraints $C$ (such as
number of connections of a given graph) \cite{FerreriCanchoSole2003a, ColizzaRinaldo2004,
Newman2010}. A similar example in another field models languages as a network of associations
between objects and words, and considers language evolution through a least effort process
\cite{FerreriCanchoSole2003b, ProkopenkoPolani2010, SalgeProkopenko2013}. Here, the cost-efficiency
conflict is mapped onto coding/decoding efforts for users of an economic (while ambiguous) language.
Both in the linguistic and the network examples, the tension between opposite demands leads to phase
transition phenomena. We can wonder, in a more general note, when and how will phase transitions
arise disregarding of the details of the problem in hand.

  Within the context of network optimization, we consider the set $\Gamma$ of all connected networks
$\gamma \in \Gamma$ involving $N$ nodes and any number of links. Latter on we will define an
efficiency ($E(\gamma)$) and a cost ($C(\gamma)$) based on the structure of each network $\gamma$,
so that a series of optimization problems can be posed. For each such graph, we can also introduce a
global energy function $\Omega(\gamma)$ that takes into account our optimization goals. The most
straightforward way to do this is through a linear combination:
    \begin{equation}
      \Omega(\gamma, \lambda) = \lambda E (\gamma) + (1 - \lambda) C(\gamma), 
      \label{eq:1}
    \end{equation} 
with $\lambda \in [0,1]$ a tunable parameter that weights the impact of each contribution. This is
precisely the strategy in previous accounts of these problems \cite{FerreriCanchoSole2003a,
ColizzaRinaldo2004, Newman2010, FerreriCanchoSole2003b, ProkopenkoPolani2010, SalgeProkopenko2013,
MathiasGopal2001}. If $\lambda=1$ only efficiency constraints will be at work, whereas $\lambda=0$
would ignore this component.

\begin{figure}
        \begin{center}
          \includegraphics[width=8 cm]{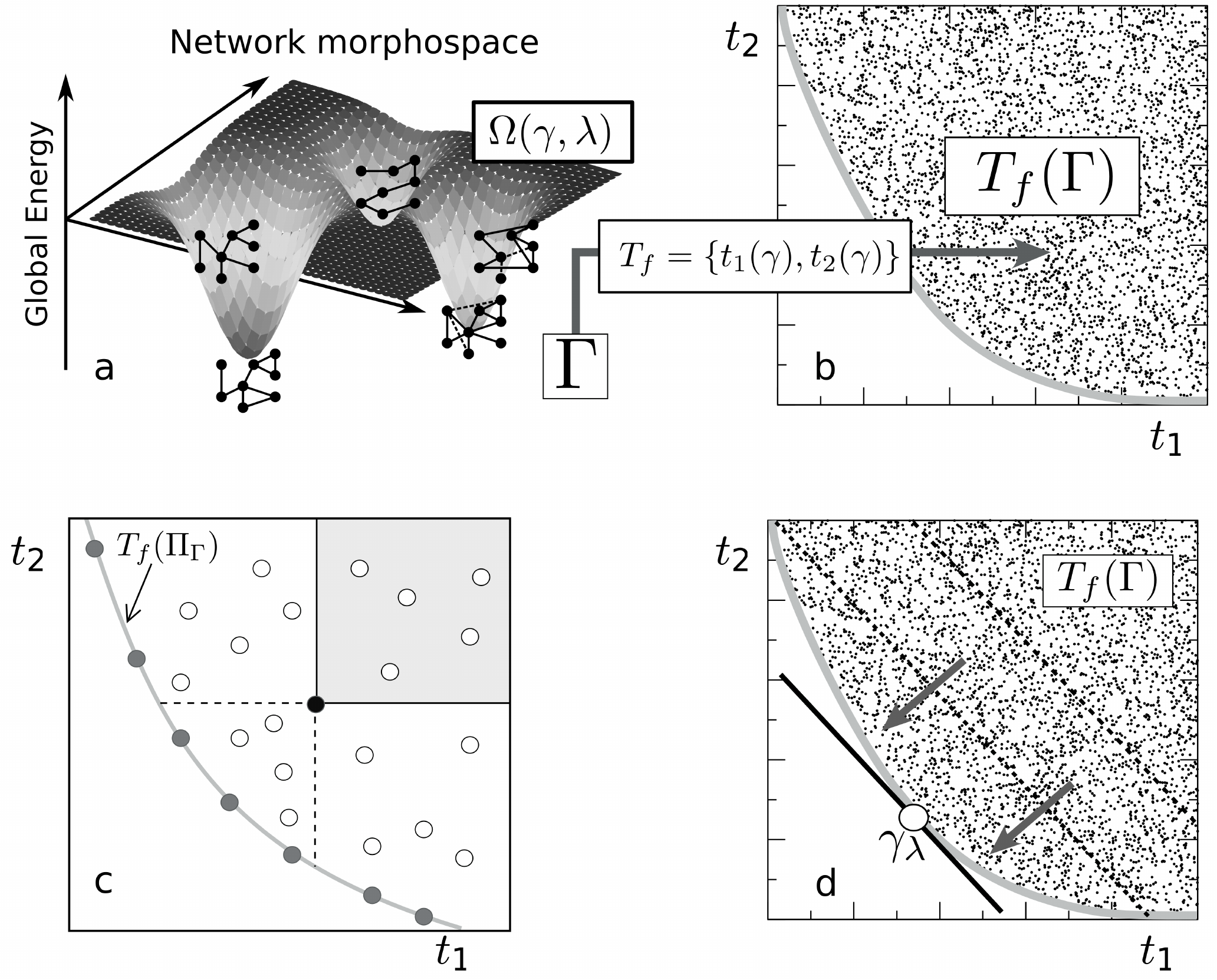}

          \caption{\textbf{A two dimensional example of Pareto optimality. } (a) $\gamma \in \Gamma$
are all possible connected networks with a given number of nodes. They populate some network
morphospace where we seek those graphs minimizing some measurable feature. If we deal with just one
fitness function, an energy landscape can be defined and the optima are easily found at the bottom
of energy wells. (b) If more than one optimization target are at play, this landscape picture falls
apart and we need to adopt a Pareto optimization approach. Then our task is to find a set of Pareto
optimal solutions ($\Pi_\Gamma \subset \Gamma$) that minimizes all targets (here $t_1$ and $t_2$)
simultaneously. These functions map each network $\gamma \in \Gamma$ into $\mathbb{R}^2$. The subset
of Pareto optimal solutions is mapped into the Pareto front (thick gray curve). Along this curve it
is not possible to improve both $t_1$ and $t_2$ at the same time. (c) It is convenient to introduce
{\em Pareto dominance}. The black circle represents a net with lower $t_1$ and $t_2$ than (thus it
Pareto dominates) those within the gray square. That same filled circle is dominated by those
networks projected between the dashed lines and the Pareto front. (d) Recovering a single objective
optimization through a linear combination of the targets (Eq. (\ref{eq:1})) is akin to choosing a
direction in $\mathbb{R}^2$ and seeking the extreme of the front along that direction. }

          \label{fig:1}
        \end{center}
      \end{figure}

  Such a global energy function results in very illustrative visualizations of the optimization
process through the notion of a {\em potential landscape}. Assuming minimization, more optimal
network architectures lay deeper in a potential well when we plot $\Omega(\gamma, \lambda)$ for
every $\gamma$ superimposed on an arbitrary network morphospace, as in Fig. \ref{fig:1}(a). Note,
however, that this is a limited picture: a fixed value of $\lambda$ is necessary to generate one
potential landscape. Changing the parameter modifies the landscape rendered by $\Omega(\gamma,
\lambda)$ and, accordingly, the underlaying optimization problem. To achieve a more general
comprehension we should not only allow scenarios with different values of $\lambda$, but we must
also question the hypothesis of linearity introduced by equation (\ref{eq:1}). Therefore, we
consider Pareto (or Multi Objective) Optimization \cite{FonsecaFleming1995, Coello2006,
Schuster2012}, whose solution is not a global optimizer dependent on external parameters (i.e. the
absolute minimum of some potential landscape), but a collection of solutions that attempt to satisfy
both (cost and efficiency) constraints simultaneously. These Pareto optimal designs constitute the
so-called Pareto front: the most optimal tradeoff possible between the targets involved (Fig.
\ref{fig:1}(c)).

  In this paper we look at such tradeoffs for a series of optimization problems defined upon complex
networks. How the different constraints are satisfied depending on a series of factors is
interesting in itself; but if we relate the Pareto optimal designs to the linear optimization
problem posed by equation (\ref{eq:1}), a richer phenomenology unfolds. This stems from a deep
connection between the Pareto front and phase transitions and other important features of
thermodynamic theory \cite{SeoaneSole2013}. In this context, the energy landscapes that we can
generate through equation (\ref{eq:1}) often result very useful to complete the picture.

  The paper is organized as follows. Sec. \ref{sec:2} provides the basic formalism of Pareto
optimization and the connection between Pareto optimality and statistical mechanics
\cite{SeoaneSole2013}. This theory homes in phase transitions for the problems investigated and is
the framework used to analyze most of the results obtained. Also in Sec. \ref{sec:2} we pose a
series of Multi Objective Optimizations upon complex networks whose solutions are detailed in Sec.
\ref{sec:3} and discussed, with closing remarks, in Sec. \ref{sec:4}. Appendix \ref{app:1} explains
some numerical aspects of the current work.

\section{Phase transitions in the Pareto formalism}
 \label{sec:2}

  A recent contribution proposed a multiobjective optimization approach to statistical mechanics
that generalizes key concepts in thermodynamics for any Pareto optimality problem
\cite{SeoaneSole2013}. The fundamental connection is between the Pareto front and the Gibbs surface.
This surface (defined through the thermodynamic potential $G=G(U,S,V)$) is linked to the equilibrium
state of a thermodynamic species. Then, its concavities and non-differentiable edges underly the
existence of first and second order phase transitions \cite{Gibbs1873, Maxwell1904}.

  A similar point can be made for systems that optimize a set $T_f$ of target functions $T_f = \{
t_1, ..., t_K \}$, whose Pareto front encodes phase transitions and critical points in its shape.
This connection between Pareto optimal systems and thermodynamics is explained at length in
\cite{SeoaneSole2013}. There, general Pareto optimal designs (not necessarily networks) that
minimize an arbitrary number $K \ge 2$ of targets is discussed. Here we sketch the theory with just
$K=2$ optimization targets and using connected networks with fixed number of nodes ($\gamma \in
\Gamma$). This suffices to illustrate the relevant aspects of the theoretical framework. We remit
the reader to \cite{SeoaneSole2013} for further details.

  Henceforth, $T_f = \{t_1, t_2 \}$ are any two real valued functions that we can measure on any
network. These functions project each network into $T_f(\gamma) = (t_1(\gamma), t_2(\gamma))$, a
point in the plane that we term {\em target space} (Fig. \ref{fig:1}(a)-(b)). In that plane we can
solve the Multi Objective Optimization (MOO) problem consisting of the simultaneous minimization of
both targets. Therefore we define Pareto dominance (Fig. \ref{fig:1}(c)): A network $\gamma_x \in
\Gamma$ dominates another $\gamma_y \in \Gamma$ (and we note it $\gamma_x \prec \gamma_y$) if
    \begin{eqnarray}
      t_k(\gamma_x) &\le& t_k(\gamma_y) \>\> \forall \>\> k=1,2; \nonumber \\ 
      \exists \> k'\in\{1,2\} &|& t_{k'}(\gamma_x) < t_{k'}(\gamma_y). 
    \end{eqnarray}
In this case $\gamma_x$ is objectively better than $\gamma_y$ and we can dismiss the later. It is
often the case that pairs of networks are mutually non dominated. Then we cannot choose between them
without introducing a bias towards either $t_1$ or $t_2$. We have to avoid this bias to find Pareto
optimal designs. We say that a network $\gamma_x \in \Gamma$ is Pareto optimal if it does not exist
any other $\gamma_y \in \Gamma$ such that $\gamma_y \prec \gamma_x$. Hence
    \begin{equation}
      \Pi_\Gamma = \{ \gamma_x \in \Gamma \;\; \vert \;\; \nexists \; \gamma_y \in \Gamma, \; \gamma_y\prec \gamma_x \}
    \end{equation} 
is the set of all Pareto optimal solutions, which constitutes the solution of the MOO problem. 

    The set $\Pi_\Gamma \subset \Gamma$ is projected onto the target plane through $T_f(\Pi_\Gamma)$
where it represents a limiting frontier of the whole $T_f(\Gamma)$, known as the {\em Pareto front}.
For $K=2$ this frontier is a curve that implements a bijective function of $t_1$ and $t_2$ (Fig.
\ref{fig:1}(c)). $\Pi_\Gamma$ is not a standard global optimizer. Instead, it comprises a collection
of valid networks that embody the optimal tradeoff between many targets such that, as we move
through it, we cannot improve $t_1$ without worsening $t_2$ and vice-versa.

  We can now add a further demand that a global energy function 
    \begin{eqnarray}
      \Omega(\gamma; \lambda) &=& \lambda t_1(\gamma) + (1-\lambda)t_2(\gamma)
      \label{eq:5}
    \end{eqnarray}
be minimized. (This equation is equivalent to Eq. \ref{eq:1} with the more general $t_{1,2}(\gamma)$
instead of efficiency and cost. We rewrite it here for convenience.) Note that this is the simplest
Single Objective Optimization (SOO) that we can built with $T_f$. Indeed, each possible $\lambda \in
[0,1]$ poses a different SOO problem, potentially with a distinct global solution. We check now how
solutions of this SOO family are related.

  By setting fix a value of $\lambda$ we choose a direction in the $t_1 - t_2$ plane along which the
minimization proceeds (Fig. \ref{fig:1}(d)). From Eq. \ref{eq:5}, networks projected onto a straight
line with slope $d = -\lambda / (1-\lambda)$ have the same global energy $\Omega$. Networks on
straight lines {\em pushed} further against the front present lower energies, until one solution
$\gamma_{\lambda} \in \Pi_\Gamma$ is singled out as global minimum of $\Omega(\lambda)$. It usually
lays where the slope of the front is precisely $d = -\lambda / (1-\lambda)$. If the front
$\Pi_\Gamma$ is convex and its derivatives are well defined, we can smoothly sample the front by
slowly tuning $\lambda$. Any appropriate measurement that we perform on $\gamma_\lambda$ will be a
smooth, differentiable function of $\lambda$ itself.

      \begin{figure}[htbp]
        \begin{center}
          \includegraphics[width=8 cm]{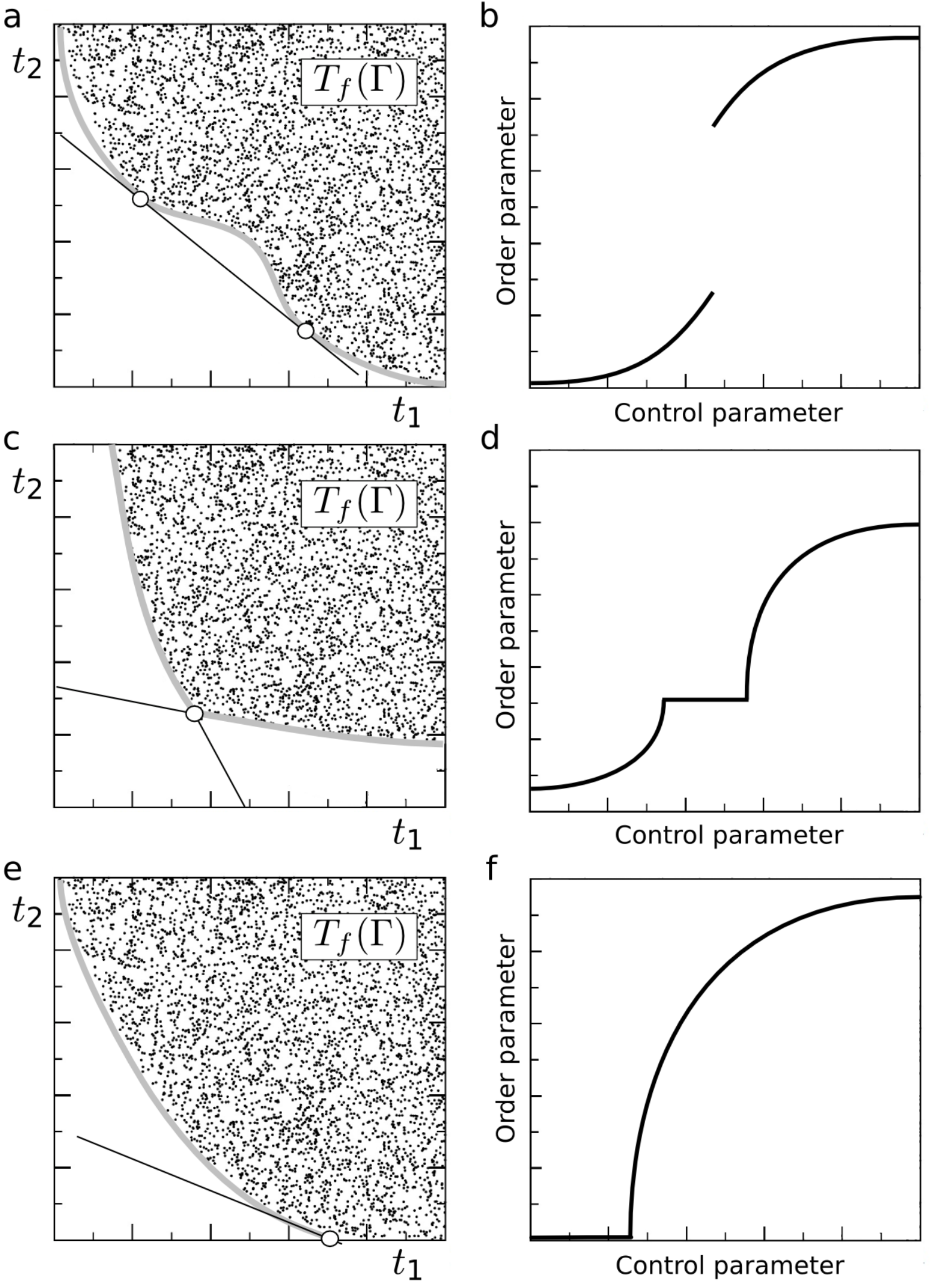}

          \caption{\textbf{Phase transitions in Pareto optimal systems. } (a) A first order phase
transition takes place in MOOs whose Pareto front presents a cavity. Solutions of the convex hull of
the front are SOO global optima for a range of $\lambda$, while those inside the cavity are not. (b)
This renders a gap in any adequate, measurable property of the global solutions. (c) Second order
phase transition are associated to ill-defined derivatives -- i.e. sharp edges in the front. Those
designs at the edge are persistently optimal foe a range of the control parameter $\lambda$. (d)
Hence, any property of the global solution is constant as a function of that same control parameter
for the designated interval. (e) Second order phase transitions might also take place at the
extremes of the Pareto front. We say then that the front has an abrupt ending (or that it ends in an
abrupt manner) as in the lower-right extreme of this front. (f) Any order parameter presents the
well-known functional dependence of second order phase transitions. }

          \label{fig:2}
        \end{center}
      \end{figure}

  If a cavity exists in $T_f(\Pi_\Gamma)$, the global solutions bypass those that lay inside (Fig.
\ref{fig:2}(a)). For a certain value $\lambda^c$, two very different solutions
($\gamma^1_{\lambda^c}$ and $\gamma^2_{\lambda^c}$) located far apart in the $t_1 - t_2$ plane are
simultaneously optimal. Solutions between $\gamma^1_{\lambda^c}$ and $\gamma^2_{\lambda^c}$ (inside
the cavity) never get to be global optima. For $\lambda < \lambda^c$ we remain at one side of the
cavity where a smooth sample of the Pareto front is still possible. The same happens for $\lambda >
\lambda^c$ at the other side of the cavity. But at the characteristic value $\lambda^c$ there is a
sudden jump between $\gamma_{\lambda^c}^1$ and $\gamma_{\lambda^c}^2$ and, consequently, in any
order parameter that we can measure on $\gamma_{\lambda}$ when plotted as a function of $\lambda$
(Fig. \ref{fig:2}(b)). This is a first order phase transition. A similar transition happens if the
Pareto front is a straight line (then $\lambda^c = -d^c / (1 - \lambda^c)$, where $d^c$ is the slope
of the front) with the singularity that all Pareto optimal solutions are also global optima for
$\lambda = \lambda^c$.

  If the front is convex but its slope is ill-defined anywhere (Fig. \ref{fig:2}(c)) there is a
range of values $\lambda \in (\lambda^-, \lambda^+)$ for which the global optimum remains the same.
Because the optimum does not change within this singular range, if we plot any order parameter its
derivative (with respect to $\lambda$) will be zero. The same derivative will be different from zero
anywhere else. This results in a characteristic plot for any order parameter (Fig. \ref{fig:2}(d))
associated to second order phase transitions.

  Such non-differentiability of the front might be shifted all the way towards one of its extremes.
For $\lambda \in (0,1)$ we seek global optima where the slope of the front is any $-\lambda /
(1-\lambda) \Rightarrow d \in (-\infty, 0)$. Well-behaved Pareto fronts should roll smoothly with
slopes that completely span the $d \in (-\infty, 0)$ interval; otherwise the front will be exhausted
while more SOO problems can still be defined. If these slope values are not exhausted we say that
the Pareto front {\em ends in an abrupt manner}, meaning that its slope as we approach its leftmost
end is some $d^*_- > -\infty$, or its slope as we approach its rightmost end is some other $d^*_+ <
0$. These abrupt endings of the front will be optima already for some $\lambda^{*} = -d^*_{\pm} /
(1-d^*_{\pm})$. We can still define SOO problems for $\lambda$ beyond these limits, for which the
solutions at the corresponding extremes of the front will be persistently optimal (Fig.
\ref{fig:2}(e)). This implies again a discontinuity in the derivative of any order parameter, the
fingerprint of second order transitions (Fig. \ref{fig:2}(f)).

  \subsection{Multiobjective optimization of complex networks}
    \label{sec:2.1}

    Complex graphs are a great testbed to illustrate this theoretical framework. They allow us to
define problems of increasing difficulty where first and second order transitions arise. Seminal
work on network optimization addressed the problem from an SOO perspective \cite{GastnerNewman2006,
FerreriCanchoSole2003a, Newman2010, MathiasGopal2001, LoufBarthelemy2013}, so some of our results
can be put in context. Another advantage of complex networks is that good optimizers can be produced
in the computer, simplifying the empirical work.

    We propose three problems based on the conflict between the average path length between nodes
and the density of edges, which roughly inform us about diffusion efficiency \cite{GoniSporns2013}
(for which low average path length is desired) and implementation costs (lower for sparser
networks). Consider first the {\em topological} (or standard) average path length:
      \begin{eqnarray}
        \left< l \right>^{t}(\gamma) &=& {1 \over Z^t_{\left< l \right>}} \sum_{i,j} {d^t_{ij}(\gamma) \over 2}, 
        \label{eq:app.01}
      \end{eqnarray}
where $d^t_{ij}(\gamma)$ denotes the distance (in number of edges) between nodes $n_i, n_j \in
\gamma$ along the shortest path that connects them; and the topological (or standard) link density:
      \begin{eqnarray}
        \rho^t(\gamma) &=& {1 \over Z^t_{\rho}} \sum_{i,j} {a_{ij}(\gamma) \over 2}, 
        \label{eq:app.03}
      \end{eqnarray} 
where the adjacency matrix $A(\gamma) = \{a_{ij}(\gamma)\}$ presents $a_{ij}(\gamma) = 1$ if two
nodes are linked in $\gamma$ and $a_{ij}(\gamma) = 0$ otherwise. $Z^t_{\left< l \right>}$ and
$Z^t_\rho$ are normalization constants discussed below.

  The superindices in $\left< l \right>^{t}(\gamma)$, $\rho^t(\gamma)$ indicate that we deal with
the topological (or standard) average path length and link density, in which edges cost $1$ unit.
Geometric costs can be included if nodes are distributed, e.g., over a Euclidean space. Let
$d^g_{ij}(\gamma)$ be the Euclidean length of the shortest path connecting $n_i$ and $n_j$ in
network $\gamma$ -- i.e. the sum of the Euclidean lengths of the edges in the shortest path between
these nodes provided that $\gamma$ is embedded in some geometric space. We introduce the geometric
(or weighted) average path length:
            \begin{eqnarray}
        \left< l \right>^{g}(\gamma) &=& {1 \over Z^g_{\left< l \right>}} \sum_{i,j} {d^g_{ij}(\gamma)  \over 2}. 
        \label{eq:app.02}
      \end{eqnarray}
The shortest Euclidean distance possible between two nodes $l_{ij}(\gamma)$ only enters equation
(\ref{eq:app.02}) if a direct link between $n_i$ and $n_j$ is present in $\gamma$ (in that case
$d^g_{ij}(\gamma) = l_{ij}(\gamma)$). This $l_{ij}(\gamma)$ allows us to introduce the geometric (or
weighted) link density:
      \begin{eqnarray}
        \rho^g(\gamma) &=& {1 \over Z^g_{\rho} } \sum_{i,j} {a_{ij}(\gamma)l_{ij}(\gamma) \over 2}. 
        \label{eq:app.04}
      \end{eqnarray}

    Just as before, $\left< l \right>^{g}(\gamma)$ and $\rho^g(\gamma)$ (note the superindexes
indicating their geometric dependence) are normalized by $Z^g_{\left< l \right>}$ and $Z^g_{\rho}$.
A clique, or fully connected network ($\gamma_C$), has the shortest average path length possible
always. As we will see later, this means that $\gamma_C$ is Pareto optimal {\em always}, so we base
our normalization on it: $Z^{t/g}_{\left< l \right>} = \sum_{i,j} d^g(i,j; \gamma_C)/2$, $Z^t_{\rho}
= \sum_{i,j} a_{ij}(\gamma_C)/2$, and $Z^g_{\rho} = \sum_{i,j} a_{ij}(\gamma_C)l(i,j;
\gamma_C)/2$.\\

  We combine $\left< l \right>^{t/g}(\gamma)$ and $\rho^{t/g}(\gamma)$ as targets in different ways
to generate three MOO problems:
    \begin{itemize}

      \item[(A)] Fully topological problem, with $t_1 = \left< l \right>^t(\gamma)$ and $t_2 =
\rho^t(\gamma)$. Note that the geometry does not play any role in this case.

      This version was originally studied in \cite{FerreriCanchoSole2003a, Newman2010} from an SOO
perspective. From that approach, only the clique and star graphs appear relevant (as discussed in
\cite{Newman2010}) as the representatives of two phases at either side of a discontinuous phase
transition. We show how this fits parsimoniously within the framework presented in
\cite{SeoaneSole2013}. But besides, we discuss now the whole Pareto front -- its relevant shape and
some of its constituents. This front includes non-trivial complexities well differentiated from the
star and clique, and it presents connections to critical systems discussed elsewhere
\cite{SeoaneSole2015}.

      \item[(B)] Partly geometrical problem,  with $t_1 = \left< l \right>^t(\gamma)$  (the same as
above) and $t_2 = \rho^g(\gamma)$. Geometry, through $t_2$, plays a relevant role now. Since the
disposition of the nodes in space matters, we study this MOO in two different cases: i) nodes
scattered randomly over the $[0,1]\times[0,1]$ square in $\mathbb{R}^2$ and ii) nodes spaced evenly
over a circle of radius $1$.

      In this problem we still use the topological average path length, meaning that we seek to
minimize the number of hops or the number of relay stations between arbitrary pairs of nodes. To
think about this problem we can picture an infrastructure such as a subway network whose contractor
wishes to minimize the length of line built while the users want to avoid transfers between lines.

      \item[(C)] Fully geometrical problem, with $t_2 = \left< l \right>^g(\gamma)$ and $t_2 =
\rho^g(\gamma)$. In this case the geometrical cost is important for all targets involved. Again, the
disposition of the nodes matters and again we study: i) nodes scattered randomly over the
$[0,1]\times[0,1]$ square in $\mathbb{R}^2$ and ii) nodes spaced evenly over a circle of radius $1$.

    \end{itemize}

\section{Results}
	\label{sec:3}

  Three relevant topologies indicate major feats of all our Pareto fronts. The most prominent one is
the clique: a fully connected network that presents the largest number of links possible, thus
maximizes edge density and minimizes the average path length always. This guarantees that the clique
is always Pareto optimal. It marks the top-left boundary of the Pareto front, as illustrated in Fig.
\ref{fig:3}(a). This is true for all problems considered in this paper.

  The star presents a hub to which all other nodes are connected, while non-hubs are not connected
to each other. There are $N$ possible star graphs. If geometry is not considered, all of them are
equivalent. When geometry intervenes and nodes are spaced over a circle all $N$ stars are equivalent
as well. All possible trees consist of as many edges as the star but, if geometry matters, only the
minimum spanning tree (MST) minimizes {\em always} the edge density ($t_2$). The MST is Pareto
optimal whenever geometry is relevant and it always indicates the end of the Pareto front opposite
to the clique (at its bottom-right).

	\subsection{Fully topological problem}
		\label{sec:3.01}

      \begin{figure}[htbp]
        \begin{center}
          \includegraphics[width=8 cm]{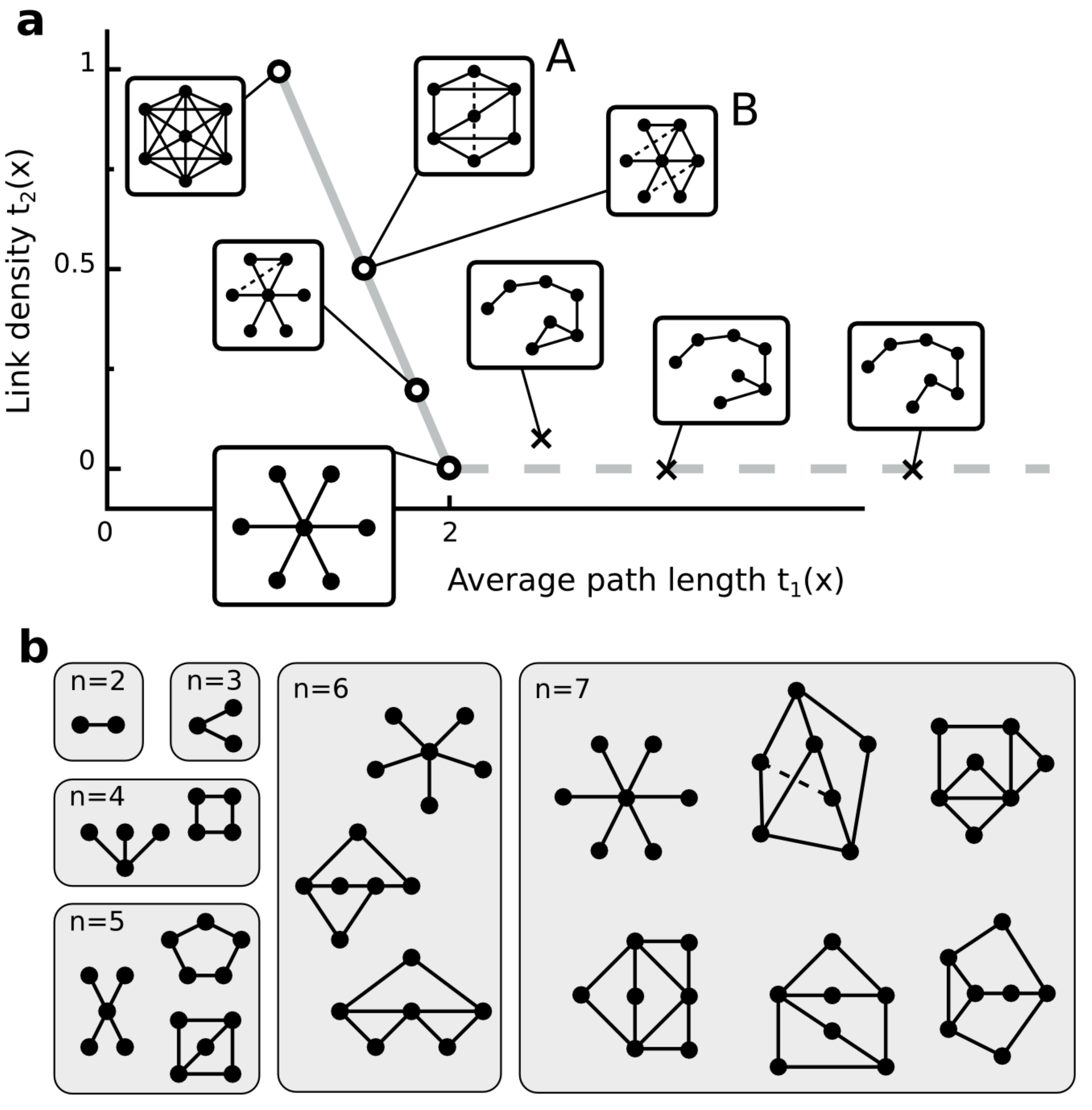}

          \caption{\textbf{Pareto front of the fully topological problem. } (a) The front (solid
gray curve) is a straight line connecting two phases: a star and a clique. The slope of the line
$d^c = -1$ determines that at $\lambda^c = 1/2$ a first order phase transition takes place. All
networks laying on the front are global SOO optima at that critical value. Among them we find
networks produced by attaching links to a star and others radically different from the star and from
the clique (note the two graphs marked A and B: only one of them can be produced by attaching edges
to the star). (b) All {\em core graphs} for have been listed for $N \le 5$. Beyond that, it becomes
increasingly difficult to count how many there are or even to tell apart two different ones. }

          \label{fig:3}
        \end{center}
      \end{figure}

    This case has been studied as an SOO through equation (\ref{eq:5}) \cite{FerreriCanchoSole2003a,
Newman2010}. That solution is incomplete from an MOO perspective which was not the chosen paradigm
in those works anyway. This problem has the advantage that its front (Fig. \ref{fig:3}(a)) can be
found analytically and the phase transitions derived from it are independent of the number of nodes.
We cannot guarantee the same for the variations studied later.

      \begin{figure*}[htbp]
        \begin{center}
          \includegraphics[width=16 cm]{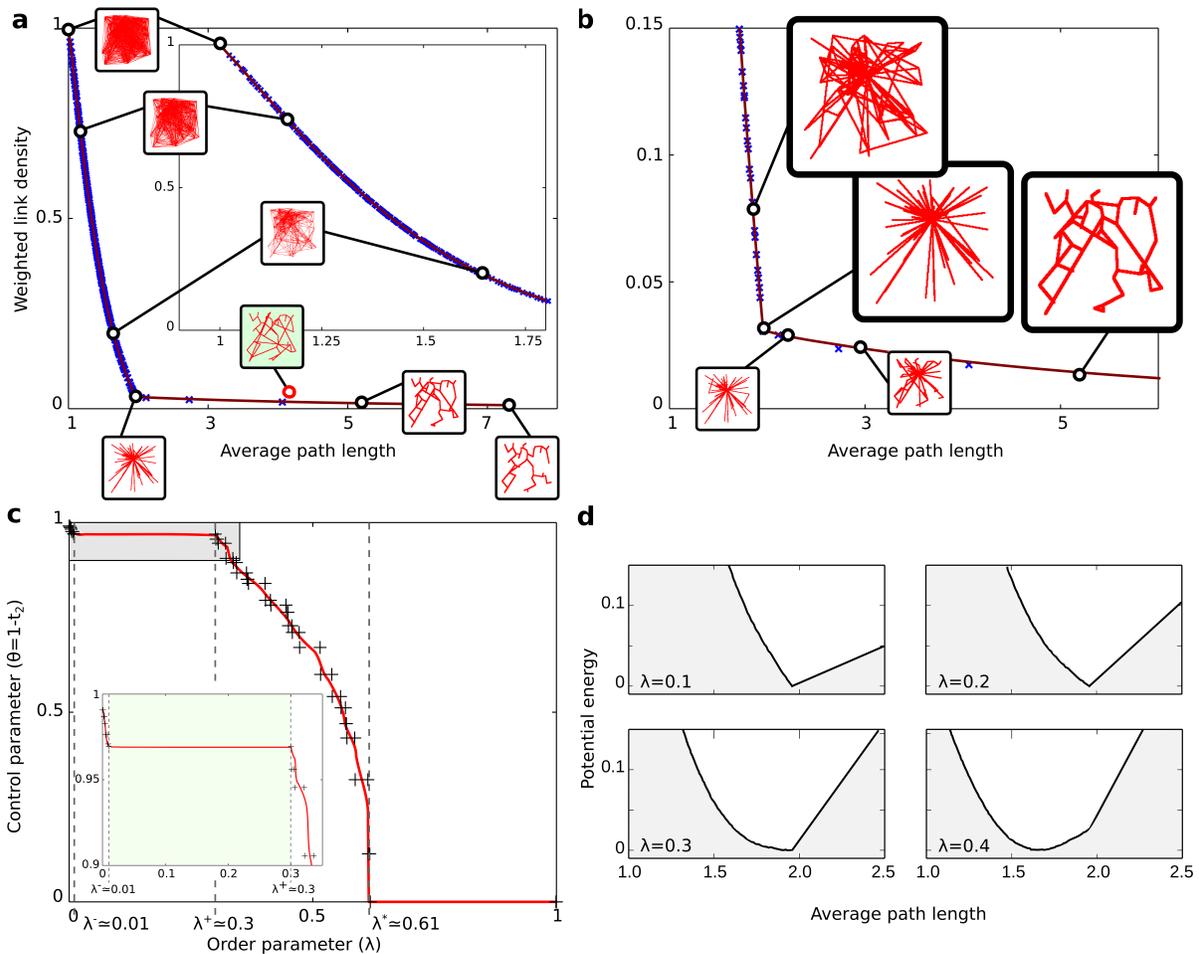}

          \caption{\textbf{Partly geometrical problem on nodes scattered over a plane. } (a) The
front follows the archetype of the topological problem with two roughly perpendicular stretches that
trade off between the clique (top-left), the star, and the MST (bottom-right). Incomplete cliques
are reached after a second order phase transition because the Pareto front ends abruptly in its
top-left (inset). The other extreme of the front ends smoothly. (b) A sharp edge indicates a second
order phase transition with the star graph being optimal for a range $\lambda \in (\lambda^-,
\lambda^+)$. (c) Plotting an order parameter as a function of $\lambda$ reveals both transitions at
$\lambda \simeq 0.61$ and at $\lambda^- \simeq 0.01$ and $\lambda^+ \simeq 0.3$ (inset). The star is
optimal in the range $\lambda \in (\lambda^-, \lambda^+)$, thus any order parameter is constant in
that range. (d) The non-analyticity of the Pareto front is inherited by the energetic landscape also
as a sharp edge. The SOO is vividly illustrated thanks to this potential landscape, whose minimum is
occupied by one same network for several values of $\lambda$. As lambda changes, the potential well
is deformed until the minimum drifts away from the sharp edge. }

          \label{fig:4}
        \end{center}
      \end{figure*}

      Because we normalized both targets using the clique as a reference, this network is mapped
into $(1, 1)$ in the $t_1 - t_2$ plane. Any graph will have less edges than a clique, thus the set
$\Gamma$ of all connected networks lays below $t_2 = 1$ in the target plane. The lower boundary of
$t_2$ is achieved by connected networks with the minimum amount of edges possible ($N - 1$). There
are a collection of such graphs, from the star to a linear chain -- in between lay all possible
trees. All them have $t_2 = \rho^t = 1/N$, which tends to $0$ as $N$ goes to infinity. The average
path length of these networks varies between that of the star ($2 (N-1) / N \rightarrow 2$) and the
linear chain ($(N^2-1)/3(N-1) \rightarrow +\infty$). These minimally connected graphs lay on a
horizontal stretch of the $t_1 - t_2$ plane (dashed line in Fig. \ref{fig:3}(a)).

      Among these trees (all with the same $t_2 = 1/N$), the star is the one with the lowest average
path length, hence it is Pareto optimal. Any other network with a lower $t_1$ must have more links
than the star, the clique setting the lower $t_2$ bound. Thus the Pareto front must lay on a curve
connecting the clique and the star -- i.e. connecting $(1, 1)$ and $(2, 0)$ in the $t_1 - t_2$
plane. 

      We appreciate the following facts: i) The edge density is a function of the number of links
alone and it does not depend on the topology of the network. ii) Given a network that is Pareto
optimal, we generate new Pareto optimal networks by simply adding new connections. As an instance,
the star is Pareto optimal and all its nodes are $1$ edge apart from the hub and $2$ edges apart
from each other. Then, new edges can only be added that connect directly two non-hub nodes, turning
a distance $d^t(i,j)=2$ into $d^t(i,j)=1$; but not affecting the network in any other respect. Put
otherwise, once a network is Pareto optimal any addition of links has got only {\em local} effects
in its average path length. 

      Adding new links to the star results in more Pareto optimal networks, the number of which
grows combinatorially (that scaling saturates as we approach the clique). Take apart the $N - 1$
non-hub nodes of a star: any network that we implement on this subset of nodes (connected or not),
and which is subsequently embedded on the original star graph through the hub, is Pareto optimal. It
is a {\em sufficient} (but not necessary) condition for a network to be Pareto optimal to contain a
hub (Fig. \ref{fig:3}(a)). The {\em necessary} condition for Pareto optimality is that every node is
at maximum $2$ edges apart from each other. 

      From any Pareto optimal network (with or without a hub), adding new edges always generates new
Pareto optimal graphs. Repeating this operation we always reach a clique, but this process is not
reversible: Take the clique and delete connections randomly with the condition that your network
remains Pareto optimal after every deletion. No rearrangement of the edges is allowed. Let this
process continue until we cannot remove any link without violating the Pareto-optimality condition.
This algorithm might yield a star or any other graph from a collection of {\em irreducible} Pareto
optimal networks, which we call {\em Pareto core graphs}. The star is the core graph with less edges
possible. We can only construct these networks as described, since other defining regularities are
not apparent -- beyond the optimality condition that every node is at most $2$ hops away from each
other. Some of these graphs are represented in Fig. \ref{fig:3}(b) for $N = 2, \dots, 7$. The
complexity scales from $1$ core graph for $N = 2, 3$; to two core graphs for $N = 4$; to three for
$N=5$; to an unknown number for $N \ge 6$. For larger $N$ it also becomes increasingly difficult to
determine whether two core graphs are the same, given their invariance under the labeling of the
nodes. Note that core graphs {\em are} Pareto optimal. They are representative of the staggering
complexity contained in the front (which grows combinatorially) and they cannot be trivially
composed as a mixture of stars and cliques. Because of this they constitute Pareto and global optima
that have not been previously reported.

      Even if we cannot list down all Pareto optimal networks, we can find where they live on the
$t_1 - t_2$ plane. Adding one edge always modifies $\left< l \right>^t(\gamma)$ by an amount $\Delta
\left< l \right>^t = -1/N(N-1)$, thus $t_1$ is decreased. The same operation increases $t_2$ by
$\Delta \rho^t = 1/N(N-1)$. Because $\Delta \rho^t / \Delta \left< l \right> = -1$ does not depend
on the number of edges, Pareto optimal graphs thus generated lay on a straight line with slope $d^c
\equiv \Delta \rho^t / \Delta \left< l \right>$ (Fig. \ref{fig:3}(a)). Such a front implies a first
order phase transition at $\lambda^c = -d^c/(1-d^c) = 1/2$. The clique and the star are found at
either phase (correspondingly for $\lambda > \lambda^c$ and $\lambda < \lambda^c$). Right at the
critical value $\lambda^c$ any Pareto optimal network is a global optimum. The degenerated
complexity at $\lambda = \lambda^c$, with so many and structurally different optimal solutions,
presents interesting connections with critical phenomena and neutral theory that we explore in
future work \cite{SeoaneSole2015}. The plot of any order parameter as a function of $\lambda$ (not
shown) just presents a gap between two constant values.

  \subsection{Partly geometrical problem} 
    \label{sec:3.02}

    Figure \ref{fig:3}(a) provides an archetype for the Pareto front that will be repeated (with
variations) in the more elaborated MOO problems. Our fronts will present a first, stepped stretch
that trades off between the clique (top-left) and some intermediate networks (usually the star); and
a second, flat stretch with little variation in the vertical dimension ($t_2$) and a broad variation
in the horizontal axis ($t_1$). In the previous case, this second stretch (dashed line in Fig.
\ref{fig:3}(a)) does not belong to the front, but it will in the following problems.

    \subsubsection{Nodes scattered over a plane}
      \label{sec:3.02.01}

      \begin{figure*}[htbp]
        \begin{center}
          \includegraphics[width=16 cm]{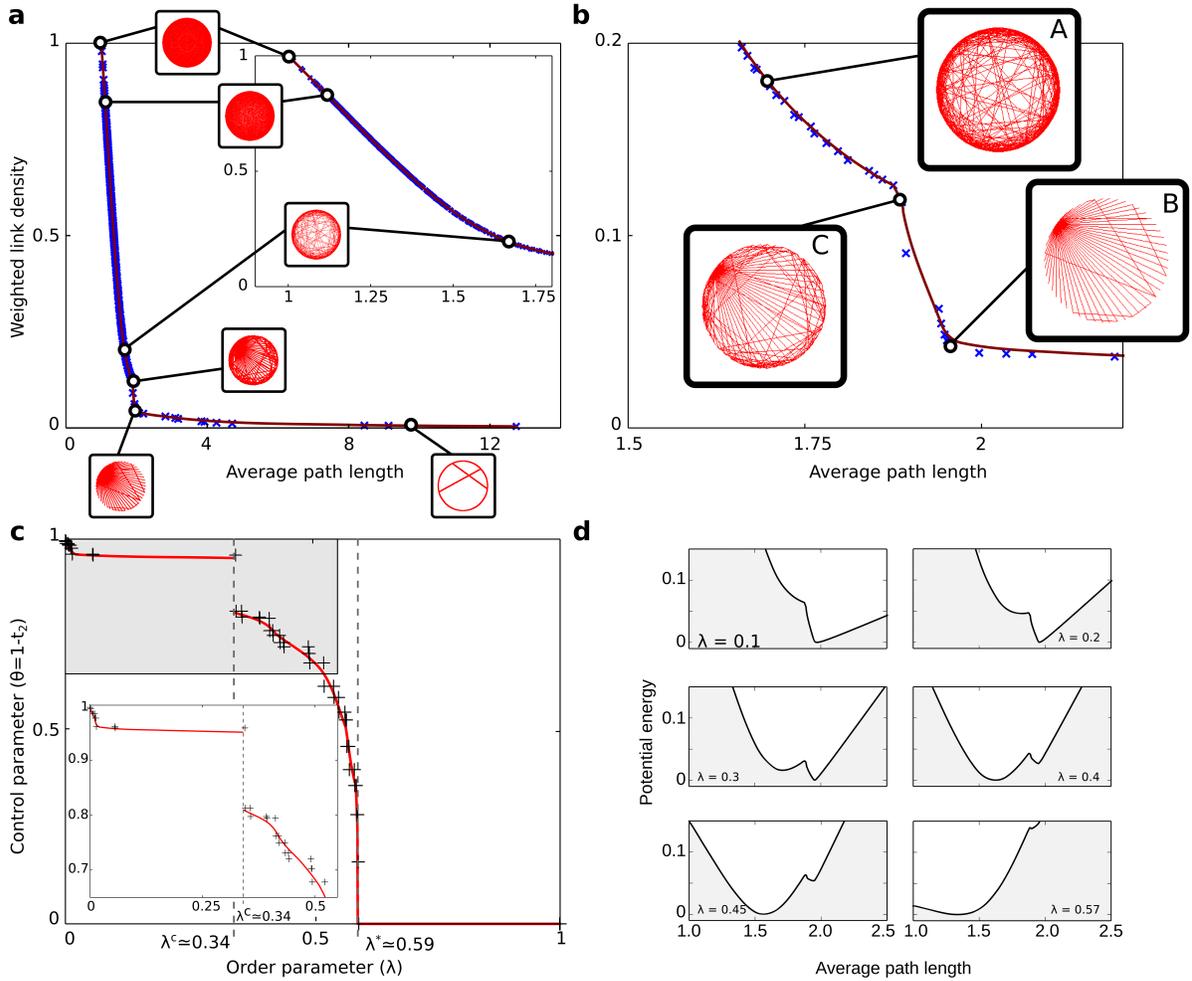}

          \caption{\textbf{Partly geometrical problem on a circle. } (a) Again, the front follows
the archetype of the topological problem with two roughly perpendicular stretches that trade off
between the clique (top-left), star networks, and the MST. Around the clique it is observed the same
phase transition as before. In the cavity it is solved a complex rearrangement. Networks that drop
their larger connections first (A) must morph into a star (B), which requires some of these  
far-reaching edges. Therefore, some Pareto optimal networks are produced that never get to be SOO
optima (C), yielding a first order transition. (c) Order parameters as a function of $\lambda$
reveal the first ($\lambda^c \simeq 0.34$, magnified in the inset) and second ($\lambda^* \simeq
0.59$) order transitions. (d) The landscape potential unveils the mechanisms for local equilibrium
and hysteresis associated to first order transitions. At low levels of the control parameter
($\lambda \sim 0.1$) only one minimum exists in the global energy $\Omega$. A pocket becomes locally
stable for $\lambda \sim 0.2$. This grows for larger $\lambda$, until it becomes the global extreme
of the energetic landscape ($\lambda \sim 0.4$). Optimizing our networks through numerical
algorithms can get us stuck in local minimums, so to transit from one potential well to the other we
need to increase our control parameter until one of the wells get destabilized ($\lambda \sim
0.57$). Repeating the operation with decreasing $\lambda$ can get us stuck in the other well, thus
engaging in a hysteresis loop. Some Pareto optimal networks inside the cavity are reached at these
metastable states. }

          \label{fig:5}
        \end{center}
      \end{figure*}

      Figure \ref{fig:4}(a) shows the first example of this archetype. A very stepped stretch of the
front trades off between the clique and the star just as before. However, this is a convex curve
now, which we discuss below. The second archetypal stretch of the front trades off between the star
and the MST, and is mapped onto an almost horizontal curve in the $t_1-t_2$ plane with a slight
convexity. This Pareto front ends up smoothly in its bottom-right extreme, so we dismiss any
phenomenon associated to it. Because the whole front is convex first order phase transitions are
ruled out.

      The first, stepped stretch of the front (Fig. \ref{fig:4}(a), inset) presents a feature that
appears in most subsequent cases. This stretch is a convex curve that ends abruptly (with the notion
of {\em abruptness} introduced before). This indicates that a second order phase transition takes
place. This transition trades off between the clique (persistent global optimum for $\lambda >
\lambda^* \simeq 0.61$) and dense but incomplete graphs reached as we move below $\lambda^*$.
Because the clique is optimal for $\lambda \ge \lambda^*$, anything that we measure on this global
optimum stays constant as a function of $\lambda$ until $\lambda < \lambda^*$, for which our
wandering over the Pareto front yields a changing global optimum as $\lambda$ decreases. Then, any
measurement performed on the SOO optimum will vary steadily with a derivative (with respect to
$\lambda$) different from $0$. This discontinuity in the derivative indicates that a second order
phase transition takes place (Fig. \ref{fig:4}(c)).

      A sharp edge in the front (Fig. \ref{fig:4}(b)) indicates yet another second order phase
transition similar to that described in Fig. \ref{fig:2}(b): The slope of the front is well defined
as we tend towards the sharp edge from the left (yielding a slope $d^+$ such that $\lambda^+ =
-d^+/(1-d^+) \simeq 0.3$), and as we tend to the sharp edge from the right (now with $d^-$ such that
$\lambda^-=0.01$). For any $\lambda \in (\lambda^-, \lambda^+)$, SOOs are well defined through
equation (\ref{eq:5}); but there are not any points of the front with a slope $-\lambda/(1-\lambda)
= d \in (d^+, d^-)$ where to locate the optimum. Instead, the same one network laying precisely at
the sharp edge is consistently optimal for this range of $\lambda$. Anything that we measure about
this optimum will remain as a function of $\lambda$ in $(\lambda^-, \lambda^+)$, but samplings of
the front run smoothly below $\lambda^-$ and above $\lambda^+$, with well defined derivatives for
any order parameter as a function of $\lambda$. Hence, two discontinuities are evident in this
derivative (Fig. \ref{fig:4}(c), inset).

      Qualitatively, browsing the front through $\lambda$ is a continuous transition from the clique
(which is a global optima for a wide range of $\lambda$), through the star graph (also a persistent
optimum for a continuum of $\lambda$), to the MST. The weighted density of edges ($t_1 = \rho^g$)
penalizes large connections first, which are dropped as we leave the clique. But enough of them
survive among Pareto optimal graphs so that the star can be reached continuously, without needing a
drastic rewiring that would leave an imprint in the order parameters. Note that these few long edges
survive because they enable a low average path length ($t_2 = \left< l \right>^t$), which is still
measured as the number of hops between nodes. Finally, to rearrange the star into the MST, the
surviving long edges are replaced by winding branches that extend visiting many nodes on their way.
Alternative strategies, like hybrid MSTs that incorporate non-essential shortcuts between far-apart
nodes, fall off the Pareto front (note the sub-optimal graph highlighted in Fig. \ref{fig:4}(a)).

      It is very useful to consider the potential landscape introduced by equation (\ref{eq:5}) to
stretch our intuition. For a fixed value of $\lambda$ we compute the energy $\Omega(\gamma,
\lambda)$ for every Pareto optimal network. The result is a lower energy boundary (Fig.
\ref{fig:4}(d)). Not Pareto-optimal networks must present yet higher energies. The SOO solution
becomes now very intuitive since the global optimum lays at the minimum of $\Omega(\gamma,
\lambda)$, which has got a vivid graphic representation. But this potential landscape changes as a
function of $\lambda$, and consequently its minimum. The sharp edge of the Pareto front is inherited
by the potential landscape, which presents a persistent minimum for a range of the control
parameter.

    \subsubsection{Nodes spaced over a circle}
      \label{sec:3.02.02}

      The front of this problem (Fig. \ref{fig:5}(a)-(b)) again follows the archetype: i) a stepped
stretch to the left that trades-off between the clique and dense (yet incomplete) graphs and ii) a
flatter stretch that encompasses graphs with roughly the same edge density but a wide variation
along the average path length dimension. This second stretch extends to large values of $t_1$: a
region populated by minimal, circlelike networks with little long-range connections. It seems a
convex stretch that ends smoothly, so nothing remarkable happens there. Again, the stepped stretch
of the front is convex (Fig. \ref{fig:5}(a), inset) and ends abruptly revealing a second order phase
transition. It is similar to the one encountered before by the clique and happens at a similar value
$\lambda \simeq 0.59$ (Fig. \ref{fig:5} (c)).

      The notable feature that this front introduces is a concavity at the junction between the two
archetypal stretches of the front (Fig. \ref{fig:4}(b)). The cavity lays at the confluence between
the three relevant network topologies: incomplete cliques, the star, and encircled nets (the MST of
the problem). As in the previous case, longer edges are dropped first. The scattered nodes managed
to retain enough long-range links in that example, enabling a continuous transition through the
star. This is not possible now, and the symmetry of the circle might be crucial therefore. Earlier,
the distribution of lengths were varied, while now all long-range edges are the same: the moment one
is dropped, the others follow. As we leave the clique, we converge quickly to encircled graphs with
little long range connections; and these lay inside a cavity of the front (Fig. \ref{fig:4}(b)). To
reconstruct a star (which remains Pareto optimal due to its low average path length) a drastic
rewiring in unavoidable. This prompts a first order phase transition ($\lambda^c \simeq 0.34$) whose
imprint is, indeed, that cavity. That transition is reflected in any order parameter $\theta$ that
we plot as a function of $\lambda$ (Fig. \ref{fig:4}(c), inset).

      We can resort again to a potential landscape to visualize this transition. Plotting
$\Omega(\gamma, \lambda)$ for all Pareto optimal networks we obtain the lower energy boundaries
portrayed in Fig. \ref{fig:5}(d). This landscape changes as $\lambda$ varies, producing two
potential wells associated to local minimums. At $\lambda = \lambda^c$, both minimums present the
same energy, thus both phases coexist. Moving away from the transition point, one of the wells is
unstabilized. Note that moving $\lambda$ back and forth could get us temporarily trapped in
metastable states (the most energetic local minimum) and hysteresis loops would be observed.

	\subsection{Fully geometric problem}
		\label{sec:3.03}

    Introducing geometry in both target functions has the effect of smoothing the Pareto front,
removing first order phase transitions. Some relevant second order transitions disappear. Others
persist, but only at the extremes of the front. The picture becomes closer to a soft trading-off
between the clique and the MST.

    \subsubsection{Nodes scattered over a plane}
      \label{sec:3.03.01}

      \begin{figure}[htbp]
        \begin{center}
          \includegraphics[width=\columnwidth]{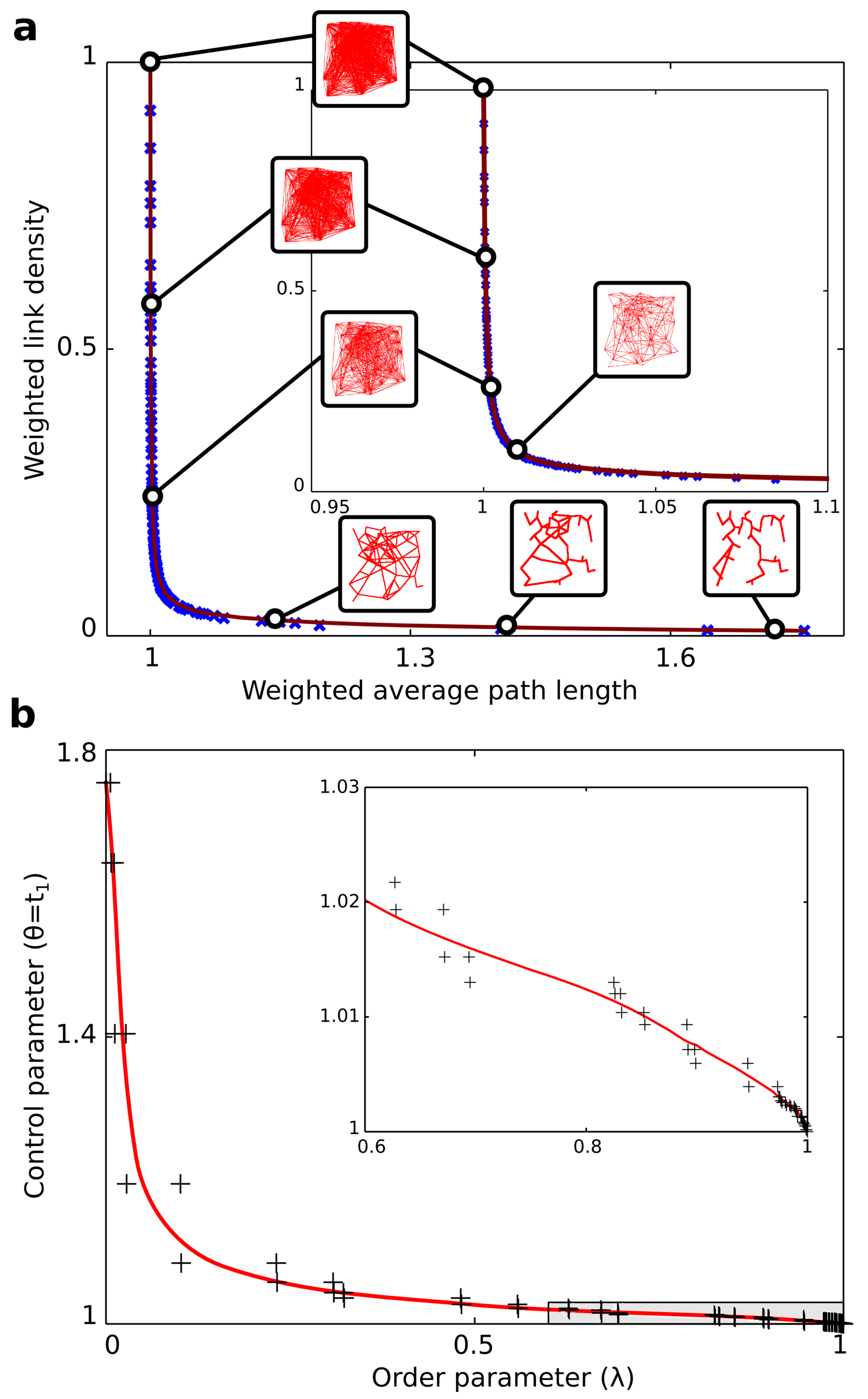}

          \caption{\textbf{Fully geometrical problem for nodes scattered over a plane. } (a) The
front has no accidents. It is completely convex and spans all possible slopes so that each $\lambda
\in (0, 1)$ poses an SOO with a different solution. As we roll over the front, the clique gently
leads to less connected networks, towards the MST. (b) The absence of phase transitions renders
smooth plots of any order parameters. }

          \label{fig:6}
        \end{center}
      \end{figure}

      This problem presents a quite uninteresting front (Fig. \ref{fig:6}(a)) without phase
transitions. The front spans all possible slopes $d \in (-\infty, 0)$, so that SOOs with different
solutions can be posed for each $\lambda \in (0,1)$. Any order parameter renders a continuous plot
(Fig. \ref{fig:6}(b)) even when zooming in to tiny details (inset). The first derivative of the
order parameters also behaves properly.

      All phase transitions from previous cases have vanished. Besides the geometric disposition of
the nodes (which has some obvious influence over what transitions are present), it is notable that
choices of optimization targets exist for which previously existing transitions disappear. The fact
that, given a same set of networks $\gamma \in \Gamma$, a choice of targets erases previous
transitions implies that a gentle evolution between radically different topologies (clique and MST)
can happen, despite the drastic modifications that we might envision necessary {\em a priori}, and
despite the phase transitions that do take place on the same graphs for other choices of targets.
This stresses the role of target functions to frame phase transitions properly.

    \subsubsection{Nodes spaced over a circle}
      \label{sec:3.03.02}

      \begin{figure}[htbp]
        \begin{center}
          \includegraphics[width=\columnwidth]{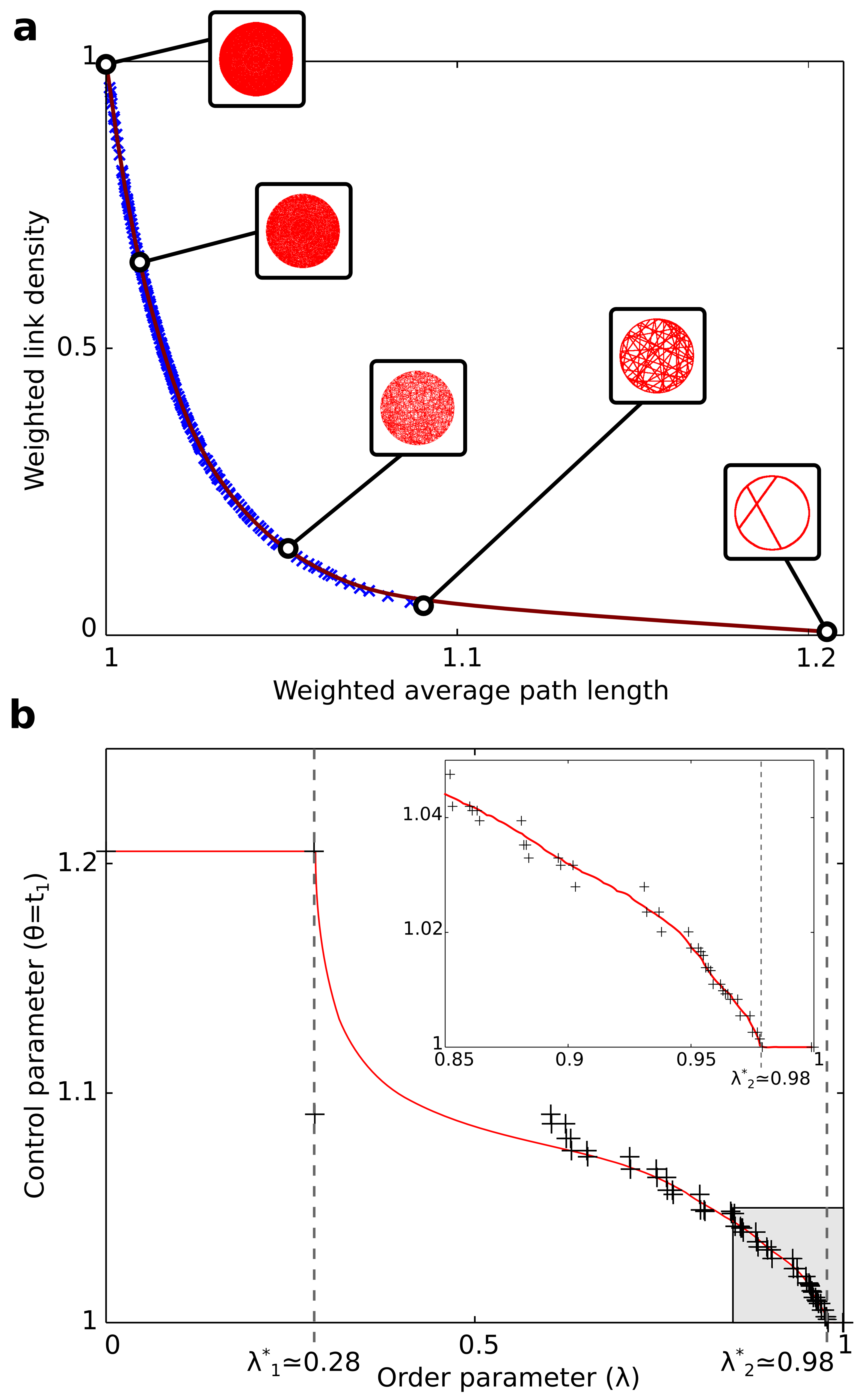}

          \caption{\textbf{Fully geometrical problem on a circle. } (a) The front presents a smooth
transition between the clique and the open circle, with no relevant feats except in the extremes of
the front. These end up abruptly, as in second order phase transitions. (b) Plotting any order
parameters reveals these phase transitions at $\lambda^*_1 \simeq 0.28$ and $\lambda^*_2 \simeq
0.98$ (inset). }

          \label{fig:7}
        \end{center}
      \end{figure}

      Again, introducing geometry in both targets smooths the Pareto front (Fig. \ref{fig:7}(a)).
The first order transition found in the circle before has disappeared. There is no cavity now and
the tradeoff between clique and circle happens gradually as we roll over the front. That previous
transition took place because the gradual drift from clique to circle was interrupted by the
presence of the Pareto optimal star, which kept the average path length low because it was measured
as the number of hops between nodes. But now geometry also enters through $t_1 = \left< l
\right>^g(\gamma)$ and using only two links to get from one node to another is still costly if these
are far reaching connections. It is more economic now to circle around even if that implies visiting
many more nodes. Thus the star is retracted from the Pareto front and the transition from the clique
to the MST proceeds smoothly.

      The bottom stretch of the front seems convex but abruptly terminated, suggesting a second
order transition at $\lambda^*_1 \simeq 0.28$ that trades off between the MST (an almost complete
circle, which is persistently optimal for $\lambda \le \lambda^*_1$) and other, more connected
graphs. The characteristic plot of a second order transition is noted in any order parameter (Fig.
\ref{fig:7}(b)). At the other end of the front we find the usual transition associated to the
clique, which did not disappear but has been moved to $\lambda^*_2 \simeq 0.98$. The same
characteristic order parameter plot can be appreciated (Fig. \ref{fig:7}(b), inset). For $\lambda
\in (\lambda^*_1, \lambda^*_2)$ any order parameter is a smooth function of $\lambda$.

\section{Conclusions}
	\label{sec:4}

  In this paper we solve three MOOs defined on complex networks. These problems allow us to explore
interesting aspects of Pareto optimality. Following recent contributions in biology
\cite{daFontouraCostaTaraskin2010, AvenaKoenigsbergerSporns2015}, our work is an exploration of a
morphospace. A first approach to such spaces is to list all possible morphologies for a system and
locate them quantitatively with respect to some relevant aspects -- here, complex networks are
characterized in an average-path-length vs. edge-density two-dimensional space. We propose that a
natural selection process based on Pareto optimality shall constrain further such a morphospace and
we study the effects of these conflictive restrictions. In doing so, we follow recent works
\cite{ShovalAlon2012, SchuetzSauer2012, SheftelAlon2013, SzekelyAlon2013} that illustrate how Pareto
optimality reduces the effective dimensionality of certain complex systems.

  Alternatively, a recent theoretical framework brings together statistical mechanics and MOO
\cite{SeoaneSole2013}, thus enriching the analysis of Pareto optimal designs. This framework reveals
universal features of Pareto optimal systems that correspond to phase transitions or critical
phenomena.

  First order phase transitions indicate that a system must undergo important structural changes
despite little variation of some control parameter. In thermodynamics, this implies great
investments of energy in reshaping matter, e.g., as it transits from solid to fluid. Similar demands
might be requested of complex Pareto optimal systems, specially if the parameters controlling the
conditions for optimality may change over time. This underscores the importance of gathering
knowledge about the Pareto front before implementing solutions to any optimization problem. 

  Second order phase transitions can also be very informative about the nature of a system. They
indicate that some solution is stable for a large range of the control parameters, thus making it
more likely if evolution or design has taken place under many different scenarios. On the other
hand, if such stable solutions would not show up in an evolutionary setup, we would have strong
evidence that a large set of possible circumstances do not occur naturally.

  We have found a variety of first and second order transitions. These depend very much on the
precise mathematical expression of the optimization targets (see, for example, how transitions
disappear as we change the measure of average path length). Hence, looking at the problem from an
alternative perspective, the presence or absence of expected phase transitions in real systems could
be informative about the nature of the optimization pressures that these systems might be subjected
to.

  Finally, the fully topological problem presents a quite singular case: a first order phase
transition takes place between the star graph (for $\lambda < \lambda^c$) and the clique ($\lambda >
\lambda^c$), while all other Pareto optimal networks are also global optima at $\lambda =
\lambda^c$. This would suggest that cliques and star graphs should happen overwhelmingly more often
than any other topology when geometry is not relevant, unless every optimal network had evolved
under the unlikely condition $\lambda = \lambda^c$. This is notably at odds with the reality and a
possible solution (with connections to critical phenomena) will be explored elsewhere
\cite{SeoaneSole2015}.

\section*{Acknowledgments}

    We thank the members of the CSL for useful discussions. This work was supported by grants from
the Fundaci\'on Bot\'in, the European Research Council (ERC Advanced Grant), and by the Santa Fe
Institute.

\appendix 

  \section{Analytic and numeric approaches for MOO solving}
    \label{app:1}

    We relied on genetic algorithms to locate the different Pareto fronts. The fully topological
case can be solved analytically, but the same genetic algorithm was used to check for good
convergence showing very good results. In this appendix we explain in detail the genetic algorithm.
Following that explanation, a few disclaimers are in order about the numerical nature of the
solutions found -- i.e. about the fact that convergence to the Pareto front cannot be guaranteed and
the smoothing necessary to render a continuous approximation to the front.

    \subsection{A multiobjective genetic algorithm}
    
      \label{app:1.01}

      MOO relays on the concept of Pareto dominance. Given two networks $\gamma_i, \gamma_j \in
\Gamma$, both mapped into $\mathbb{R}^2$ through $t_1(\gamma_{i/j})$ and $t_2(\gamma_{i/j})$, we say
that $\gamma_i$ dominates $\gamma_j$ (and note it $\gamma_i \prec \gamma_j$) if $\gamma_i$ is not
worst than $\gamma_j$ regarding any target and it is better than $\gamma_j$ with respect to at least
one target. We can visualize this: Since we deal with minimizations in the $t_1 - t_2$ plane,
network $\gamma_i$ has got a set of axes associated with their origin at $(t_1(\gamma_i),
t_2(\gamma_i))$ and every network $\gamma_j$ laying on the first quadrant of these axes is Pareto
dominated by $\gamma_i$.

      Following the literature on multiobjective genetic algorithms \cite{Dittes1996, Zitzler1999,
KonakSmith2006} we computed a dominance score: We took the set $D_j \equiv \{\gamma_i | \gamma_i
\prec \gamma_j \}$ of solutions from within a given {\em population} (an arbitrary subset of
$\Gamma$) that dominate $\gamma_j$. The size of this set ($d_j=||D_j||$, dubbed the {\em dominance
score}), indicates how {\em fit} $\gamma_j$ is in terms of Pareto optimality. We proceeded then to
minimize this score. We departed from an initial population of $N_P$ networks (either random or
designed, see below), selected $N_P/2$ of the population based on the dominance score, chose random
pairs among the selected networks to produce $N_P/2$ new networks, and applied random mutations to
all but a subset of {\em elite} networks. We iterated this scheme a fixed number of generations.

      Mutations consisted in random appending or deleting edges or totally swapping the connections
of two arbitrary nodes. For crossover, from each of the two mating nets we assigned each node and
its connections randomly to each of the offspring graphs checking that the same node and connection
was not assigned twice to the same {\em child}. After crossover or edge deletion we checked that all
networks remained connected all the time. We completed one missing link whenever connectedness
failed and then checked for connectedness again.

      All we care about for the current research is good convergence towards the front. This
justifies our using of crossover: this is a very good evolutionary operator, though unrealistic if
we wanted to study some features of nature. For example, such an operator would not be adequate to
study species that do not reproduce sexually. Studying Pareto optimality under constrained
conditions -- e.g. without crossover -- also renders a set of non-dominated solutions. These might
converge to the Pareto front or not, and they might be subjected to geometrical constraints in $t_1
- t_2$ that are similar to those studied in \cite{SeoaneSole2013} and in this paper for Pareto
optimal networks. Such constrained evolutionary schemes pose interesting research questions, but
here we are concerned with Pareto optimal solutions. This justifies the crossover and a clever
initialization of the algorithm. It might be difficult to converge towards some solutions that are
highly non-trivial -- e.g. the MST. We know, though, that this solution belongs to the front of all
physically grounded problems. If an algorithm would fail in finding this solution, this could hinder
convergence towards an interesting (though challenging) region of the Pareto front. Once again,
because we are concerned with Pareto optimal solutions and would like to attain the closest
convergence possible, it is also fully justified to {\em seed} the initial population with a few
{\em designed} solutions. We did so by introducing since the very beginning cliques, star-graphs,
MSTs, and circle networks (these two are equivalent for the circle) with very slight mutations. We
produced $N_P/4$ of each such major topologies at the beginning. The crossover and mutation
operators ensure fast exploration of hybrid topologies.

      As noted above, we know that some of these networks are Pareto optimal: the clique is always
so, and the MST is Pareto optimal in all physically grounded problems. However, and since we started
with little variations upon these graphs, the algorithm did not always reach these solutions -- but
it surely explored the region nearby. There might be other interesting regions of the front that
might not have been fully explored and that are impossible to seed without foreknowledge. Although
we are concerned with Pareto optimal solutions alone, our methods are numerical at the end and
convergence of multiobjective genetic algorithms to the Pareto front cannot be guaranteed. We
decided to report on the results of the simulations with as little reinterpretation and further
speculation as possible. Notwithstanding, the overall details of the Pareto front seem to be
recovered and the theory posed in \cite{SeoaneSole2013} is properly illustrated. \\

      As for the implementation of the algorithm, we used a population of $N_P = 3000$ connected
networks with $N = 50$ nodes -- with the initial population seeded as indicated above. The
population was evolved during $T = 10000$ generations in every case. Mutation happened with a
probability $p_\mu = 0.001$ of appending an extra link to each network, the same probability of
deleting an existing edge, and the same probability of swapping the ends of each existing
connection. The top $N_e = 50$ networks of the population where considered {\em elite} and were
spared any mutation. As the algorithm proceeds, many networks reach a dominance score of $0$ even if
they are not Pareto optimal. Unluckily, this score is the best indicator of Pareto optimality
available (not only in the current implementation, but generally). This results in elite members of
the population not being objectively better than non-elite members -- in terms of Pareto optimality.
Because the algorithm sorted the population consistently from one generation to the next, what
members of the population are considered elite is largely a matter of antiquity: early members that
reach low dominance score and are not overthrown are likely to be preserved during the whole
simulation. We repeated $4$ times the simulations with scattered nodes to check that the relevant
features obtained were not artifacts of some lucky distributions of the nodes.

  \subsection{Smoothing of the front and order parameters} 
    \label{app:1.02}

    The only speculation that we allowed ourselves is in choosing a relevant scale for analysis,
which led to a smoothing of the Pareto front. As noted in Sec. \ref{sec:3}, we deal with a discrete
set of networks whose front is necessarily discrete as well. Accordingly, every shift in global
optima is a first order phase transition at some scale and global optima remain so for a continuous
range of $\lambda$, as in second order phase transitions. This does not further our understanding of
the problem as much as a coarse-grained analysis that renders noteworthy phase transitions. Since
the genetic algorithm only produces a finite set of (ideally) Pareto optimal solutions, we applied a
Bezier smoothing to their plot on the $t_1 - t_2$ plane. We took care that the smoothing did not
introduce alien concavities. Because Bezier curves cannot present sharp edges (thus ruling out
second order phase transitions), when a sharp edge seemed the best description of the front (Sec.
\ref{sec:3.02}, partly geometrical problem on nodes scattered over a plane), we decided to split the
front in its two salient branches and apply two independent smoothing processes that allowed us to
recover the transition in great detail.

    To locate global optima, we calculated $\Omega(x_{\Pi}, \lambda)$ for the optimal solutions
produced by the genetic algorithm, and for a large sample of points from the Bezier curves
introduced in the previous paragraph. We registered the global optimal for different values of
$\lambda$. One of the problems pointed at earlier is that, because of the discreteness of the front,
global optima are so for several values of $\lambda$. This would cause that the plots of order
parameters look tiered. For a better illustration of the results, whenever order parameters are
plotted we indicate only the first and last values of $\lambda$ for which each global optima are
indeed optima (black crosses in all order parameter plots). The smoothing allows a finner grained
sampling so that this is not an issue: the corresponding order parameters (red curves in all order
parameter plots) look continuous always.\\

    Following \cite{SeoaneSole2013}, anything {\em well behaved} that we measure upon global optima
are accepted as order parameters. By well behaved we imply that order parameters should not
introduce alien divergences into the problem, and that solutions laying at different points over the
front should score differently in this parameter. This way we ensure that any feature stemming from
the optimization problem does not go unreported and that we do not introduce phase-transition--like
behaviors that originated, e.g., on some function diverging to infinity for reasons of its own.
Taking these guidelines into account, the target functions themselves are always good order
parameters. We use these ($\theta = t_1$ in Sec. \ref{sec:3.03}), or trivial transformations of them
($\theta = 1-t_2$ in Sec. \ref{sec:3.02}). More drastic transformations such as $1/(2-t_1)$ would be
banned: note that this function diverges for $t_1=2$ even if this is a perfectly regular point of
the front for all problems.


\begin{thebibliography}{99}

    \bibitem{Dawkins1997}
      R. Dawkins, 
      {\em Climbing mount improbable} 
      (WW Norton \& Company, 1997). 

    \bibitem{Ozaktas1992}
      H. M. Ozaktas, 
      Paradigms of connectivity for computer circuits and networks. 
      {\em Opt. Eng.} {\bf 31}, 1536 (1992). 

    \bibitem{Chen1999}
      W. K. Chen, 
      {\em The VLSI Handbook} 
      (CRC Press, Boca Raton, Fl, 1999). 

    \bibitem{BassettBullmore2010}
      D. S. Bassett, D. L. Greenfield, A. Meyer-Lindenberg, D. R. Weinberger, S. W. Moore, and E. T. Bullmore,  
      Efficient Physical Embedding of Topologically Complex Information Processing Networks in
      Brains and Computer Circuits. 
      {\em PLoS Comput. Biol} {\bf 6}(4), e1000748 (2010). 

    \bibitem{WattsStrogatz1998}
      D. J. Watts and S. H. Strogatz, 
      Collective dynamics of `small-world' networks. 
      {\em Nature} {\bf 393}, 440 (1998). 

    \bibitem{AmaralStanley2000}
      L. A. N. Amaral, A. Scala, M. Barth\'el\'emy, and H. E. Stanley, 
      Classes of small-world networks. 
      {\em P. Nat. A. Sci.} {\bf 97}(21), 11149 (2000). 

    \bibitem{Barthelemy2011}
      M. Barth\'elemy, 
      Spatial networks.
      {\em Phys. Rep.} {\bf 499}(1), 1  (2011).

    \bibitem{Kansky1963}
      K. J. Kansky, PhD thesis, University of Chicago, 1963.
      {\em Structure of transportation networks: relationships between network geometry and regional characteristics}, 
      PhD Thesis, 

    \bibitem{Pitts1965}
      F. R. Pitts, 
      A graph theoretic approach to historical geography. 
      {\em Prof. Geogr.} {\bf 17}(5), 15 (1965). 

    \bibitem{SenManna2003}
      P. Sen, S. Dasgupta, A. Chatterjee, P. A. Sreeram, G. Mukherjee, and S. S. Manna, 
      Small-world properties of the Indian railway network.
      {\em Phys. Rev. E} {\bf 67}(3), 036106 (2003). 

    \bibitem{GastnerNewman2006}
      M. T. Gastner and M. E. J. Newman, 
      Optimal design of spatial distribution networks. 
      {\em Phys. Rev. E} {\bf 74}, 016117 (2006). 

    \bibitem{CarvalhoArrowsmith2009}
      R. Carvalho, L. Buzna, F. Bono, E. Guti\'errez, W. Just, and D. Arrowsmith, 
      Robustness of trans-European gas networks
      {\em Phys. Rev. E} {\bf 80}, 016106 (2009). 

    \bibitem{CluneLipson2013}
      J. Clune, J. B. Mouret, and H. Lipson, 
      The evolutionary origins of modularity. 
      {\em Proc. R. Soc. B}, {\bf 280}(1755), 20122863 (2013).

    \bibitem{MengitsuClune2015}
      H. Mengistu, J. Huizinga, J. B. Mouret, and J. Clune, 
      The evolutionary origins of hierarchy. 
      arXiv preprint arXiv:1505.06353 (2015). 

    \bibitem{Murray1926}
      C. D. Murray, 
      The physiological principle of minimum work. I. The vascular system and the cost of blood volume. 
      {\em Proc. Nat. Acad. Sci.} {\bf 12}(3), 207 (1926). 

    \bibitem{WestEnquist1997}
      G. B. West, J. H. Brown, and B. J. Enquist, 
      A General Model for the Origin of Allometric Scaling Laws in Biology. 
      {\em Science} {\bf 276}, 122 (1997). 

    \bibitem{WestEnquist1999}
      G. B. West, J. H. Brown, and B. J. Enquist, 
      A general model for the structure and allometry of plant vascular systems. 
      {\em Nature} {\bf 400}, 664 (1999). 

    \bibitem{BanavarRinaldo1999}
      J. R. Banavar, A. Maritan, and A. Rinaldo, 
      Size and form in efficient transportation networks. 
      {\em Nature} {\bf 399}(6732), 130 (1999). 

    \bibitem{GafiychukLubashevsky2001}
      V. V. Gafiychuk and I. A. Lubashevsky, 
      On the Principles of the Vascular Network Branching. 
      {\em J. Theor. Biol.} {\bf 212}, 1 (2001). 

    \bibitem{CuntzSegev2007}
      H. Cuntz, A. Borst, and I. Segev, 
      Optimization principles of dendritic structure. 
      {\em Theor. Biol. Med. Model.} {\bf 4}, 21 (2007). 

    \bibitem{PerezEscuderoPolavieja2007}
      A. P\'erez-Escudero and G. G. de Polavieja, 
      Optimally wired subnetwork determines neuroanatomy of {\em Caenorhabditis elegans}. 
      {\em P. Nat. A. Sci.} {\bf 104}(43), 17180 (2007). 


    \bibitem{HasenstaubSejnowski2010}
      A. Hasenstaub, S. Otte, E. Callaway, and T. J. Sejnowski, 
      Metabolic cost as a unifying principle governing neuronal biophysics. 
      {\em P. Nat. A. Sci.} {\bf 107}(27), 12329 (2010). 

    \bibitem{CuntzHausser2010}
      H. Cuntz, F. Forstner, A. Borst, and M. H\"ausser, 
      One Rule to Grow Them All: A General Theory of Neuronal Branching and Its Practical Application. 
      {\em PLoS Comput. Biol.} {\bf 6}(8), e1000877 (2010). 

    \bibitem{WedeenTseng2012}
      V. J. Wedeen, D. L. Rosene, R. Wang, G. Dai, F. Mortazavi, P. Hagmann, J. H. Kaas, and W.-Y. I. Tseng, 
      The Geometric Structure of the Brain Fiber Pathways. 
      {\em Science} {\bf 335}, 1628 (2012). 

    \bibitem{FerreriCanchoSole2003a}
      R. Ferrer i Cancho and R. Sol\'e, 
      Optimization in complex networks. 
      In {\em Statistical Mechanics of Complex Networks} {\bf 625}, 114 (2003).

    \bibitem{Newman2010}
      M. E. J. Newman, 
      {\em Networks: an introduction} 
      (Oxford University Press, 2010), Chap. 14. 

    \bibitem{ColizzaRinaldo2004}
      V. Colizza, J. R. Banavar, A. Maritan, and A. Rinaldo, 
      Network Structures from Selection Principles. 
      {\em Phys. Rev. Lett.} {\bf 92}(19), 198701 (2004). 

    \bibitem{FerreriCanchoSole2003b}
      R. Ferrer i Cancho and R. Sol\'e, 
      Least effort and the origins of scaling in human language.
      {\em P. Nat. A. Sci.} {\bf 100}(3), 788 (2003).

    \bibitem{ProkopenkoPolani2010}
      M. Prokopenko, N. Ay, O. Obst, and D. Polani, 
      Phase transitions in least-effort communications. 
      {\em J. Stat. Mech.} {\bf 11}, P11025 (2010).

    \bibitem{SalgeProkopenko2013}
      C. Salge, N. Ay, D. Polani, and M. Prokopenko, 
      Zipf's Law: Balancing Signal Usage Cost and Communication Efficiency. 
      SFI working paper: 13-10-033 (2013).

    \bibitem{MathiasGopal2001} 
      N. Mathias and V. Gopal, 
      Small-worlds: how and why. 
      {\em Phys. Rev. E} {\bf 63}, 021117 (2001).

    \bibitem{FonsecaFleming1995}
      C. M. Fonseca and P. J. Fleming, 
      An Overview of Evolutionary Algorithms in Multiobjective Optimization. 
      {\em Evol. Comput.} {\bf 3}, 1 (1995). 

    \bibitem{Coello2006}
      C. A. Coello, 
      Evolutionary Multi-Objective Optimization: A Historical View of the Field. 
      {\em IEEE Comput. Intell. M.} {\bf 1}(1), 28 (2006). 

    \bibitem{Schuster2012}
      P. Schuster, 
      Optimization of multiple criteria. 
      {\em Complexity} {\bf 18}, 5 (2012). 

    \bibitem{SeoaneSole2013}
      L. F. Seoane and R. Sol\'e, 
      A multiobjective optimization approach to statistical mechanics. 
      http://arxiv.org/abs/1310.6372

    \bibitem{Gibbs1873}
      J. W. Gibbs, 
      A Method of Geometrical Representation of the Thermodynamic Properties of Substances by Means of Surfaces. 
      {\em Trans. Conn. Acad.} {\bf 2}, 382 (1873). 

    \bibitem{Maxwell1904}
      J. C. Maxwell, 
      {\em Theory of Heat} 
      (Longmans, Green, and Co., 1904), p. 195. 

    \bibitem{LoufBarthelemy2013}
      R. Louf, P. Jensen, and M. Barth\'elemy, 
      Emergence of hierarchy in cost-driven growth of spatial networks. 
      {\em P. Nat. A. Sci.} {\bf 110}(22), 8824 (2013).

    \bibitem{GoniSporns2013}
      J. Go\~ni, A. Avena-Koenigsberger, N. V. de Mendizabal, M. van den Heuvel, R. Betzel, and O. Sporns, 
      Exploring the morphospace of communication efficiency in complex networks. 
      {\em PLoS ONE} {\bf 8}, e58070 (2013).

    \bibitem{SeoaneSole2015}
      L. F. Seoane and R. Sol\'e, 
      Systems poised to criticality through Pareto selective forces. 
      In preparation (2015).       

    \bibitem{daFontouraCostaTaraskin2010}
      L. da Fontoura Costa, K. Zawadzki, M. Miazaki, M. P. Viana, and S. N. Taraskin, 
      Unveiling the neuromorphological space. 
      {\em Frontiers Comput. Neurosci.} {\bf 4}, 150 (2010). 

    \bibitem{AvenaKoenigsbergerSporns2015}
      A. Avena-Koenigsberger, J. Go\~ni, R. Sol\'e, and O. Sporns, 
      Network morphospace.
      {\em J. R. Soc. Interface} {\bf 12}(103), 20140881 (2015). 

    \bibitem{ShovalAlon2012}
      O. Shoval, H. Sheftel, G. Shinar, Y. Hart, O. Ramote, A. Mayo, E. Dekel, K. Kavanagh, and U. Alon, 
      Evolutionary tradeoffs, Pareto optimality, and the geometry of phenotype space.
      {\em Science} {\bf 336}(6085), 1157 (2012). 

    \bibitem{SchuetzSauer2012}
      R. Schuetz, N. Zamboni, M. Zampieri, M. Heinemann, and U. Sauer, 
      Multidimensional Optimality of Microbial Metabolism. 
      {\em Science} {\bf 336}(6081), 601 (2012). 

    \bibitem{SheftelAlon2013}
      H. Sheftel, O. Shoval, A. Mayo, and U. Alon, 
      The geometry of the Pareto front in biological phenotype space. 
      {\em Ecol. Evol.} {\bf 3}(6), 1471 (2013). 

    \bibitem{SzekelyAlon2013}
      P. Szekely, H. Sheftel, A. Mayo, and U. Alon, 
      Evolutionary tradeoffs between economy and effectiveness in biological homeostasis systems.
      {\em PLoS Comput. Biol.} {\bf 9}(8), e1003163 (2013). 

    \bibitem{Dittes1996}
      F. M. Dittes, 
      Optimization on Rugged Landscapes: A New General Purpose Monte Carlo Approach. 
      {\em Phys. Rev. Lett.} {\bf 76}(25), 4651 (1996). 

    \bibitem{Zitzler1999}      
      E. Zitzler, PhD thesis, Swiss Federal Institute of Technology Zurich, 1999. 
      Evolutionary Algorithms for Multiobjective Optimization: Methods and Applications, 

    \bibitem{KonakSmith2006}
      A. Konak, D. W. Coit, and A. E. Smith, 
      Multi-objective optimization using genetic algorithms: A tutorial. 
      {\em Reliab. Eng. Syst. Safe.} {\bf 91}(9), 992 (2006). 

  \end{thebibliography}
\end{document}